\documentclass[a4paper,twoside]{article}

\usepackage{epsfig}
\usepackage{subcaption}
\usepackage{calc}
\usepackage{amssymb}
\usepackage{amstext}
\usepackage{amsmath}
\usepackage{amsthm}
\usepackage{multicol}
\usepackage{pslatex}
\usepackage{apalike}
\usepackage{algorithm2e}
\usepackage[bottom]{footmisc}
\usepackage{SCITEPRESS}     

\begin{document}

\title{CARISMA: CAR-Integrated Service Mesh Architecture}

\author{\authorname{Kevin Klein\sup{1,2}\orcidAuthor{0000-0002-2924-4880}, Pascal Hirmer\sup{1}\orcidAuthor{0000-0002-2656-0095} and Steffen Becker\sup{2}\orcidAuthor{0000-0002-4532-1460}}
\affiliation{\sup{1}Mercedes-Benz AG, Sindelfingen, Germany\\
\{firstname.lastname\}@mercedes-benz.com}
\affiliation{\sup{2}Institute of Software Engineering, University of Stuttgart, Stuttgart, Germany\\
\{firstname.lastname\}@iste.uni-stuttgart.de}
}


\abstract{The amount of software in modern cars is increasing continuously with traditional electric/electronic (E/E) architectures reaching their limit when deploying complex applications, e.g., regarding bandwidth or computational power. To mitigate this situation, more powerful computing platforms are being employed and applications are developed as distributed applications, e.g., involving microservices. Microservices received widespread adoption and changed the way modern applications are developed. However, they also introduce additional complexity regarding inter-service communication. This has led to the emergence of service meshes, a promising approach to cope with this complexity. In this paper, we present an architecture applying the service mesh approach to automotive E/E platforms comprising multiple interlinked High-Performance Computers (HPCs). We validate the feasibility of our approach through a prototypical implementation.}

\onecolumn \maketitle \normalsize \setcounter{footnote}{0} \vfill

\section{\uppercase{Introduction}}
Nowadays, the automotive industry faces an impactful transformation towards Software-defined Vehicles (SDV), shifting the focus from hardware to software concerning the primary driver of functionality and innovation. Hence, the amount of software components in modern cars is rapidly growing, which on the one hand drives new features but on the other hand comes with a high complexity and the need for new software architectures. Traditional car architectures, which are based on a multitude of dedicated Electronic Control Units (ECUs) are not suitable anymore to support complex software-driven applications, such as autonomous driving. In newer architectures, a large number of these individual ECUs will be replaced with more powerful ones, referred to as High-Performance Computers (HPCs). They offer more computational resources and decrease the complexity of the E/E architecture in general by transitioning from a great number of individual ECUs to fewer HPCs connected to smaller ECUs for sensors and actors~\cite{Windpassinger2022}.

In contrast to traditional car architectures, mostly relying on protocols like Controller Area Network (CAN)~\cite{Johansson2005} and Local Interconnect Network (LIN)~\cite{Ruff2003} for communication, i.e., to send and receive signals, HPCs employ modern networking technologies, such as Automotive Ethernet~\cite{Matheus2021} enabling the migration to Service-Oriented Architectures (SOA)~\cite{Lawler2019} accompanied by features like plug-and-play of capabilities provided by services \cite{Kadry2022}. The software components of an application that is developed based on SOA can run distributed across the available HPCs to make full use of the available computing hardware. Through distribution and parallelization, the computing resources of a car can be used in an optimal fashion and results can be calculated more efficiently. Furthermore, the concrete distribution of the services across the different HPCs might change over time or a service might need to be deployed multiple times to enable load balancing.

In order to achieve such a distribution of applications, in cloud application development, microservices~\cite{Zimmermann2016} have been proven as a well-established pattern. However, they come with additional complexity, e.g., regarding networking and inter-service communication. To tackle these issues, service meshes have been developed, enabling developers to separate and implement any infrastructure-related concerns, e.g., service-to-service communication, in a dedicated layer~\cite{Koschel2021}.

\begin{figure}[t]
\centering
\includegraphics[width=\linewidth]{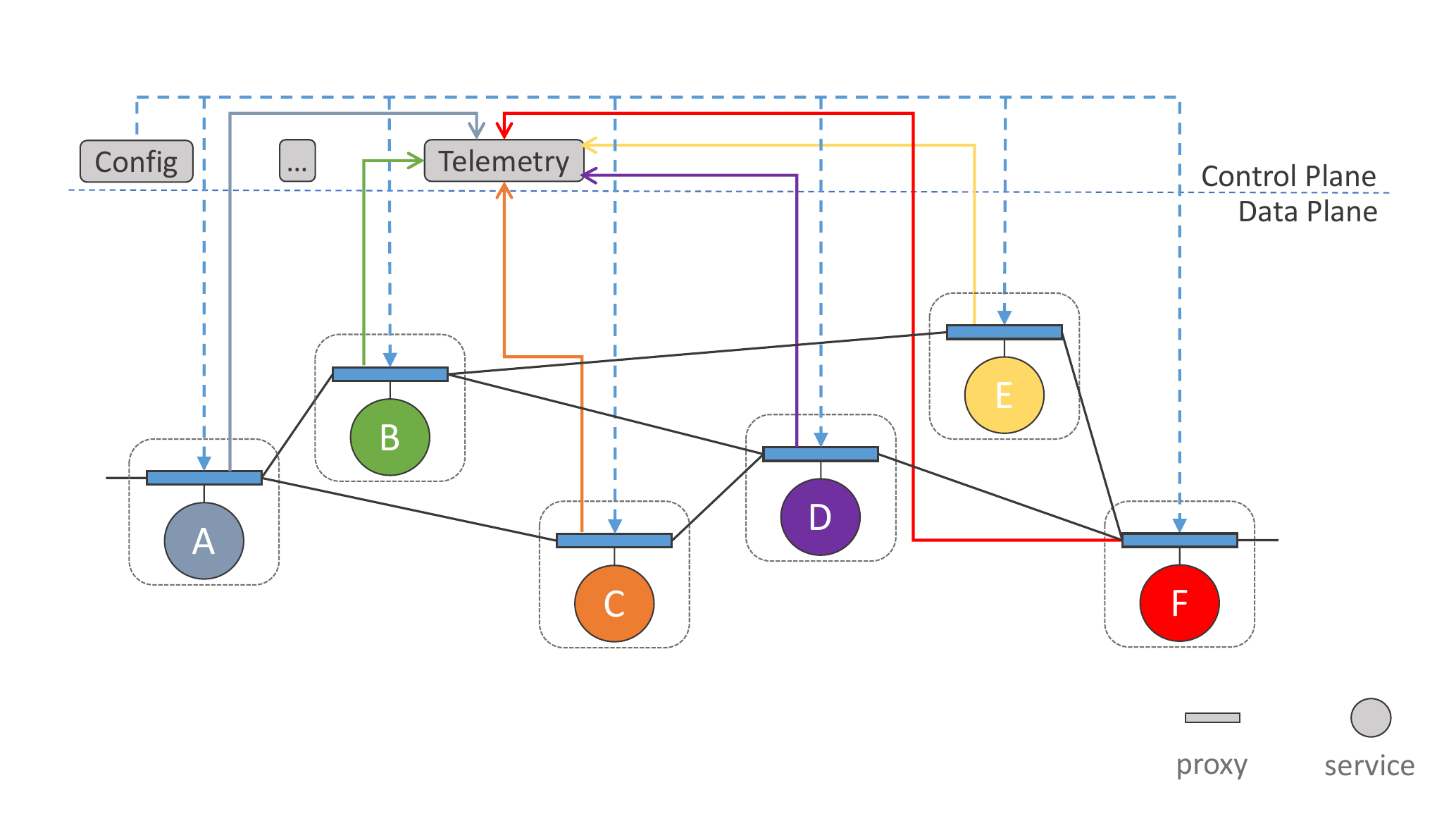}
\caption{Example applying the traditional service mesh architecture. Figure adopted with minor modifications from \cite{Li2019}.}
\label{fig:service_mesh}
\end{figure}

Figure \ref{fig:service_mesh} shows an example application that is implemented in a distributed manner with the individual services communicating through a classical service mesh. To this end, each service is accompanied by a \textit{Side-Car Proxy}, forming the \textit{Data Plane}. The proxy is responsible for routing requests from and to the service based on the configuration it receives from the configuration source that is part of the \textit{Control Plane}. Whenever a new service is deployed or re-deployed to a different location, the configuration source updates the configuration of the proxies accordingly.

Furthermore, the Control Plane has additional responsibilities, e.g., processing telemetry data collected by the proxies and forwarded accordingly.

However, the application of service meshes to modern cars also comes with a multitude of challenges. Especially, the resource limitation and limited flexibility in in-car architectures require an adaptation of the service mesh architecture to meet the specific requirements of cars. These challenges are currently not addressed by state-of-the-art approaches.

To achieve a flexible distribution of services across different HPCS, in this paper, we present CARISMA -- an in-car service mesh architecture aiming at applying the concept of service meshes that has been proven very effective in distributed cloud application development to the automotive domain. In this manner, we aim to benefit from the characteristics of service meshes to build distributed but still stable in-car applications. Furthermore, we  address the aforementioned challenges concerning the application of service meshes to modern cars. Finally, we validate the feasibility of our CARISMA approach through a prototypical validation.

\begin{figure}[t]
\centering
\includegraphics[width=\linewidth]{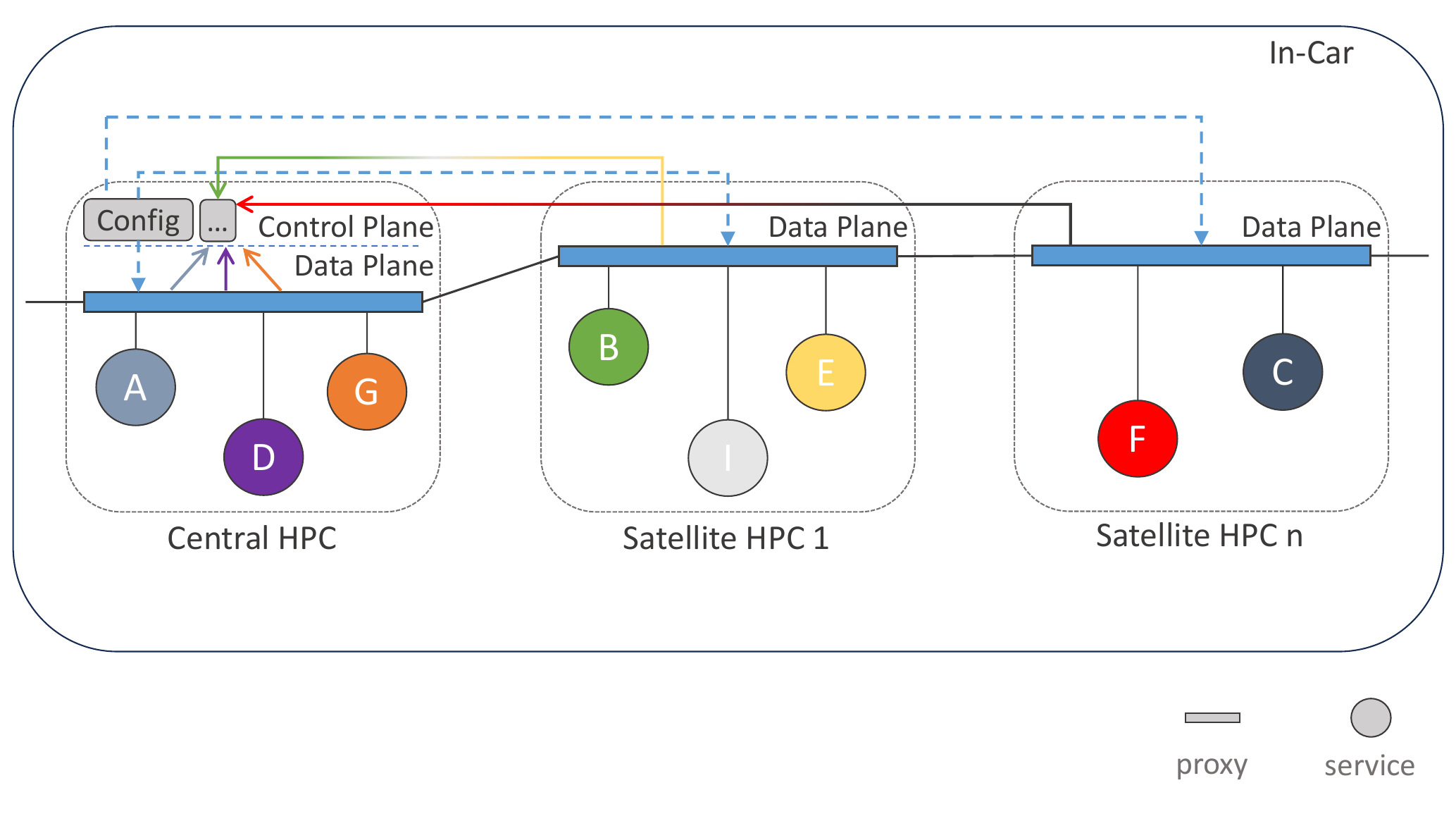}
\caption{Example applying CARISMA.}
\label{fig:carisma}
\end{figure}

Figure \ref{fig:carisma} shows an application that is designed in a distributed manner, with the individual software components being spread across three HPCs. The communication between the services is handled by our CARISMA approach. To this end, one HPC is elected as the central HPC and hosts the Control Plane. Also, we only employ one proxy per HPC that handles all incoming and outgoing service traffic for that HPC.

With the approach presented in this paper, we aim to reduce the complexity associated with the inter-service communication of distributed in-car applications. Moreover, we strive to enable a flexible re-deployment of services between individual HPCs and even load balancing of traffic between different instances of the same service independent of the concrete HPC they run on. Also, we aim to enable an easy integration of services that run within the cloud or even on edge devices.

The remainder of this paper is structured as follows: Section~\ref{sec:rw} outlines related work, whereas Sect.~\ref{sec:cm} presents CARISMA, the main contribution of this paper, followed by a description of our validation based on a prototypical implementation in Sect.~\ref{sec:pv}. Finally, in Sect.~\ref{sec:cfw}, we conclude with a summary and outline of future work.

\section{\uppercase{Related Work}}
\label{sec:rw}

In this section, we introduce related work in the scope of this paper. We examined AUTOSAR\footnote{AUTOSAR: https://autosar.org} Adaptive as well as COVESA\footnote{COVESA: https://covesa.global} since they are relevant approaches introduced by the automotive industry with AUTOSAR being the de facto standard among the big automotive vendors. Furthermore, we looked at similar approaches introduced by the research community within the field of automotive software engineering and edge computing.

Automotive Open System Architecture (AUTOSAR) is a partnership of leading automotive-related companies that defines a reference architecture for ECU software, which most manufacturers adopt. In 2017, AUTOSAR introduced a new, coexisting platform, AUTOSAR Adaptive, designed for modern automotive software with demand for high-performance computing, e.g., autonomous driving. Part of that platform is a communication management \cite{as:speccom} that provides intra-machine and inter-machine service-oriented communication between applications running on the AUTOSAR Runtime for Adaptive Applications (ARA). To this end, establishing communication between services and clients can happen statically at design time or dynamically at system start or runtime. Based on a service interface definition, a generator creates C++ classes representing a service and a client, respectively \cite{as:expcomapi}. Moreover, before services and clients can be bound dynamically, they must register with a service registry provided as part of ARA. Finding a service that is offered through the service registry is based on the generated classes and results in zero or more handles that can be used to establish the connection. In contrast, our approach does not rely on a service interface definition for establishing a communication path. Furthermore, CARISMA does not require the clients to query the service registry and initiate the service connection based on the returned handles. This enables changing the targeted service instance dynamically, e.g., to support load balancing between different service instances or to support moving a service instance from one HPC to another for optimization reasons without having the services to be aware of that change. Also, the integration of services that do not run on our platform, i.e., cloud services, is well supported by our approach.

The Connected Vehicle Systems Alliance (COVESA), formerly known under the name GENIVI Alliance, is an open development community that aims to develop open standards and technologies for connected vehicles, primarily focusing on leveraging vehicle data and vehicle-to-cloud connectivity. COVESA provides a standardized application programming interface (API) for the development of distributed, middleware-based applications named CommonAPI. Similar to AUTOSAR, a generator creates the proxy implementation and service skeleton as C++ classes based on a service interface definition file \cite{cv:userguide_commonapi}. Furthermore, some binding-specific code is generated, which is responsible for realizing the communication between a client and the service regarding a specific supported middleware technology, i.e., as of now D-Bus \cite{Love2005} or Scalable service-Oriented MiddlewarE over IP (SOME/IP) \cite{as:specsomeip}. The binding-specific code library is loaded when a proxy is created by the client application. Contrasting to CARISMA but similar to the approach chosen by AUTOSAR, this approach is based on generated code, which depends on the CommonAPI runtime library and a library containing generated binding-specific code. Consequently, breaking changes to the application binary interface (ABI) of the CommonAPI runtime library will require a recompilation of the dependent applications. Furthermore, choosing a specific service instance before creating the proxy and initiating the remote procedure call is the responsibility of the client application, making it impossible to implement load balancing strategies.

Wagner et al. \cite{Wagner2016} introduce embedded Service-Oriented Communication (eSOC), a service-oriented communication protocol for CAN. They utilize a service descriptor to encode meta information related to SOA. Since the service descriptor has a size of 64 bit and is thus quite large in comparison to the bandwidth and payload size supported by CAN, eSOC employs a short identifier that is received upon service initialization and will be used further on. That short identifier is compatible in length with the standard CAN identifier and therefore introduces no overhead. Concerning communication patterns, eSOC supports publish/subscribe as well as request/response. However, since our approach targets HPCs that are connected by a high-speed communication technology, i.e., automotive ethernet, instead of microcontrollers, an architecture built on top of CAN is not suitable.

Li et al. \cite{Li2022} propose a service mesh architecture designed to be applied in edge native computing. They argue that a plain service mesh with a Control Plane that consists of only a single controller is unsuitable for the edge because a broad distribution and a comparatively long inter-server communication delay are key characteristics. Hence, they propose to distributively deploy controllers that altogether form the Control Plane in order to avoid the single controller becoming a bottleneck. A drawback of that approach is the cost that results from keeping all the involved controllers synchronized. Consequently, they investigated how to deploy these controllers in a cost efficient way and present a customized $k$-means based algorithm for cost minimization. CARISMA, however, is designed for an environment comprising multiple HPCs that are closely located to one another. Therefore, a broad distribution and a delay in  communication are no concerns. Distributively deploying multiple controllers would introduce an unnecessary overhead resulting in higher resource consumption that can be avoided.

Furusawa et al. \cite{Furusawa2022} propose a method that employs a service mesh based on Istio\footnote{Istio: https://istio.io/} spanning multiple independent Kubernetes\footnote{Kubernetes: https://kubernetes.io} clusters to achieve a cooperative load balancing among co-located edge servers. Their goal is to reduce the loss of performance and the number of outages because of overloads. To this end, they employ a service mesh controller that monitors the number of requests for apps running on the edge servers. In case that number exceeds a certain threshold, the edge server's app is considered overloaded and another edge server with free capacities is selected. The routing is implemented using Istio's weight-based routing with the weight values to forward requests that exceed the threshold being calculated and transmitted to the Side-Car Proxies by the service mesh controller. In summary, the approach assumes an existing service mesh architecture where every app on an edge server is accompanied by a Side-Car Proxy and then implements an extension that enables an optimized distribution of the request load. In contrast, our paper focuses on implementing service meshes in a constrained environment, i.e., the automotive domain.

The presented related work is mainly focusing on implementing service-oriented architectures within the automotive domain or on a more general application of the service mesh architecture, e.g., within the context of edge computing. However, our approach is specifically designed to introduce the service mesh architecture to the automotive domain taking their specific requirements, e.g., a reduced consumption of limited resources, into account. Furthermore, our approach enables the application of methods that are known from the field of cloud application development, i.e., load-balancing and dynamic re-deployment of software components. Additionally, with CARISMA, the integration of services that are running in the cloud becomes possible without requiring them to run on top of a specific software platform.

\section{\uppercase{CARISMA}}
\label{sec:cm}

The following section presents the main contribution of this paper by introducing CARISMA -- an architecture that enables in-car applications to run distributed across a cluster of HPCs and, furthermore, to incorporate software components that run in the cloud. Figure \ref{fig:carisma} depicts an example of applying our approach, which we refer to as CAR-Integrated Service Mesh Architecture (CARISMA). It consists of the following main components: (i) a Control Plane hosting a configuration service as well as a node and service registry service storing meta-information about the nodes and services, (ii) at least one Data Plane, and (iii) exactly one service proxy per Data Plane. Compared to the traditional service mesh architecture, we decided against the Side-Car Proxy pattern, where a service proxy accompanies every service. Contrary to the cloud, for an in-car application, resource consumption is crucial. Also, the Side-Car Proxies within one HPC would not differ in terms of the configuration, making it an avoidable resource consumption. Hence, we reduced the number of service proxies to exactly one per HPC. Furthermore, we do not run the Control Plane on a dedicated node. Instead, we differentiate between a central HPC that receives a coordinative role and hosts the Control Plane next to the Data Plane and satellite nodes, which are Data Planes only. Again, the reason for this design choice is that resource consumption is crucial and we cannot afford a dedicated node running only the Control Plane due to the resource limitations within vehicles.

In a first step, every HPC has to register with the node registry service. As a result of this step, every HPC receives a unique node identifier which can then be used for service registration. As soon as a service is registered, it will become available through the service proxies. The individual steps are elaborated in the following subsections.

\subsection{Node Registration}

The Control Plane needs to be aware of the nodes, i.e. the HPCs, and their IP addresses which can be used to call services running on these nodes. To this end, we employ a node registry as part of the Control Plane. In a first step, a node announces its availability by sending a registration request containing its IP address in the request body to the node registry. The node registry maintains a mapping of a unique identifier and the IP address which the node submitted upon registration. Whenever a registration request is received, a unique identifier is generated randomly, stored together with the IP address, and is finally returned to the caller as a response to the request.

The unique identifier is used consistently across the entire configuration of the Control Plane. Therefore, it must be attached to every subsequent request, e.g., to update the service registry, via a dedicated request header. To ensure a valid configuration, the submitted unique node identifier is validated each time such a request is received.

\subsection{Service Registration}

Based on the information that is stored as part of the node registry, a per-node configuration can be generated and transmitted to the connected service proxies of the nodes. To this end, the information need to be enriched by the services that actually run on the nodes. Therefore, we furthermore employ a service registry as part of the Control Plane. Whenever a service is deployed or undeployed, the corresponding node, i.e, the HPC where the deployment or undeployment happened, has to update the service registry by sending a service registration request. Again, when sending these requests, the unique node identifier that has been received during node registration needs to be attached to the request via a dedicated request header. Whenever the service registry is updated, the Control Plane generates a new configuration snapshot and transmits it to the connected service proxies. Considering the configuration that is transmitted to the nodes, we differentiate between two configuration views:

\subsubsection{Local Configuration View}

This view maps all the services running on the same node to the IP address of the local machine and the corresponding port they are listening on.

\subsubsection{Global Configuration View} This view maps all services to the IP address of the node they are running on and the port of the \textit{ingress listener} of the corresponding service proxy.

Both views are node-specific, i.e., they do not contain the same information on every HPC. For the global configuration view, the services running on the same node, i.e., that are contained within the local configuration view, are discarded.

Every service proxy comprises two listeners: (i) the \textit{ingress listener} routing incoming traffic from other nodes to the desired local service and (ii) the \textit{egress listener} routing outgoing traffic initiated by local services to their desired target service on the same or another node. The ingress listener is configured with the local configuration view and the egress listener is configured with the local configuration view merged with the global configuration view. As described earlier, before merging the local configuration view with the global configuration view in order to attach it to the egress listener, the services contained within the local configuration view are discarded from the global configuration view. Otherwise, the exact same service would be mapped to different IP addresses resulting in an invalid configuration.

\begin{figure*}[t]
\centering
\includegraphics[scale=.3]{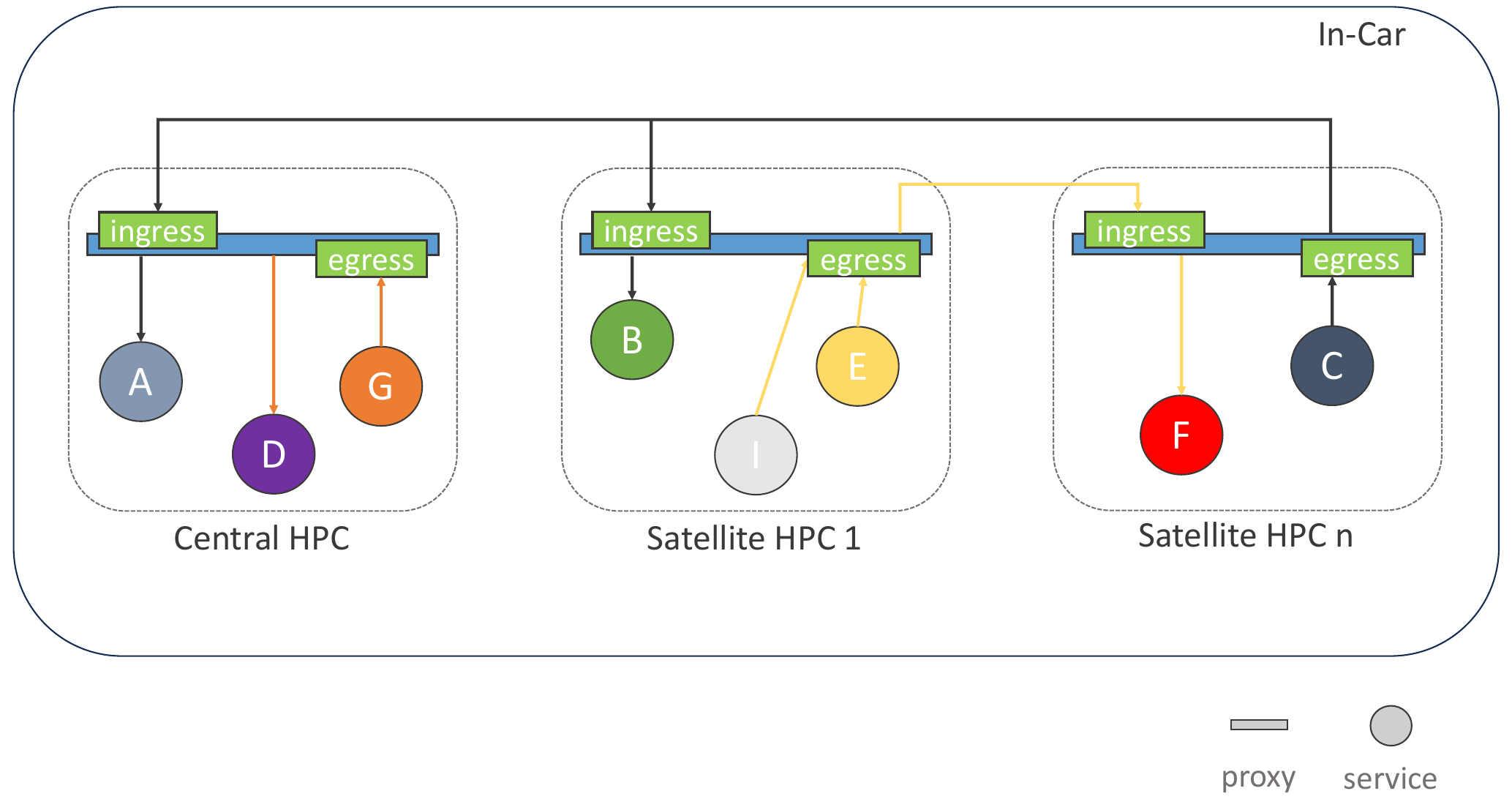}
\caption{Typical communication scenarios within CARISMA.}
\label{fig:s2s}
\end{figure*}

\subsection{Service-to-Service Communication}
 
 As soon as the nodes and services are registered, the inter-service communication becomes possible through the egress listener of the corresponding service proxy. In case the desired service is located on the same node, the request is routed to the local port on which the service listens. If, on the other hand, the service is located on a different node, the request is routed through the ingress listener of the service proxy that belongs to the corresponding target node. Figure \ref{fig:s2s} depicts typical communication scenarios: (i) communication involving services that both run on the same node (\texttt{G -> D}), (ii) communication that spans across different nodes and, in addition, is balanced between two instances of the desired target service (\texttt{A <- C -> B}), and (iii) multiple clients requesting a single service (\texttt{E -> F <- I}). Note that it is completely transparent for the caller where the desired target service is located since the communication always happens through the egress listener of the corresponding node. Even if the configuration changes and the desired target service is moved to another node, the configuration is updated instantly by the Control Plane and the caller can continue sending requests to the desired target service through the egress listener of the corresponding node.

Furthermore, CARISMA supports integrating services that run in a cloud backend in the same fashion as with services running within the HPC cluster. To this end, the endpoint that can be used to contact the cloud backend, e.g., an API gateway, needs to be registered as a node with the node registry. In consequence, it will receive a unique node identifier that can be used in subsequent requests. Furthermore, the concrete cloud service needs to be registered with the service registry. It will then be available through the egress listeners of the nodes' service proxies within the HPC cluster.

\section{\uppercase{Prototypical Validation}}
\label{sec:pv}

To validate our approach, we implemented a proof of concept application comprising the following components: (i) a Control Plane, (ii) a node registry service, (iii) a service registry service, (iv) a minimalistic orchestrator, and (v) three software components (\texttt{A, B, C}). For the implementation of (i) - (v) we relied on the Go programming language\footnote{Go Programming Language: https://go.dev} since it is widely applied in the field of cloud application development and offers great support concerning frameworks and tooling. Moreover, we chose the gRPC framework\footnote{gRPC: https://grpc.io} for communication because it is widely applied too and, furthermore, allows the implementation of an efficient and language-independent communication based on a strongly typed interface definition. 

\begin{figure*}[t]
\centering
\includegraphics[scale=.4]{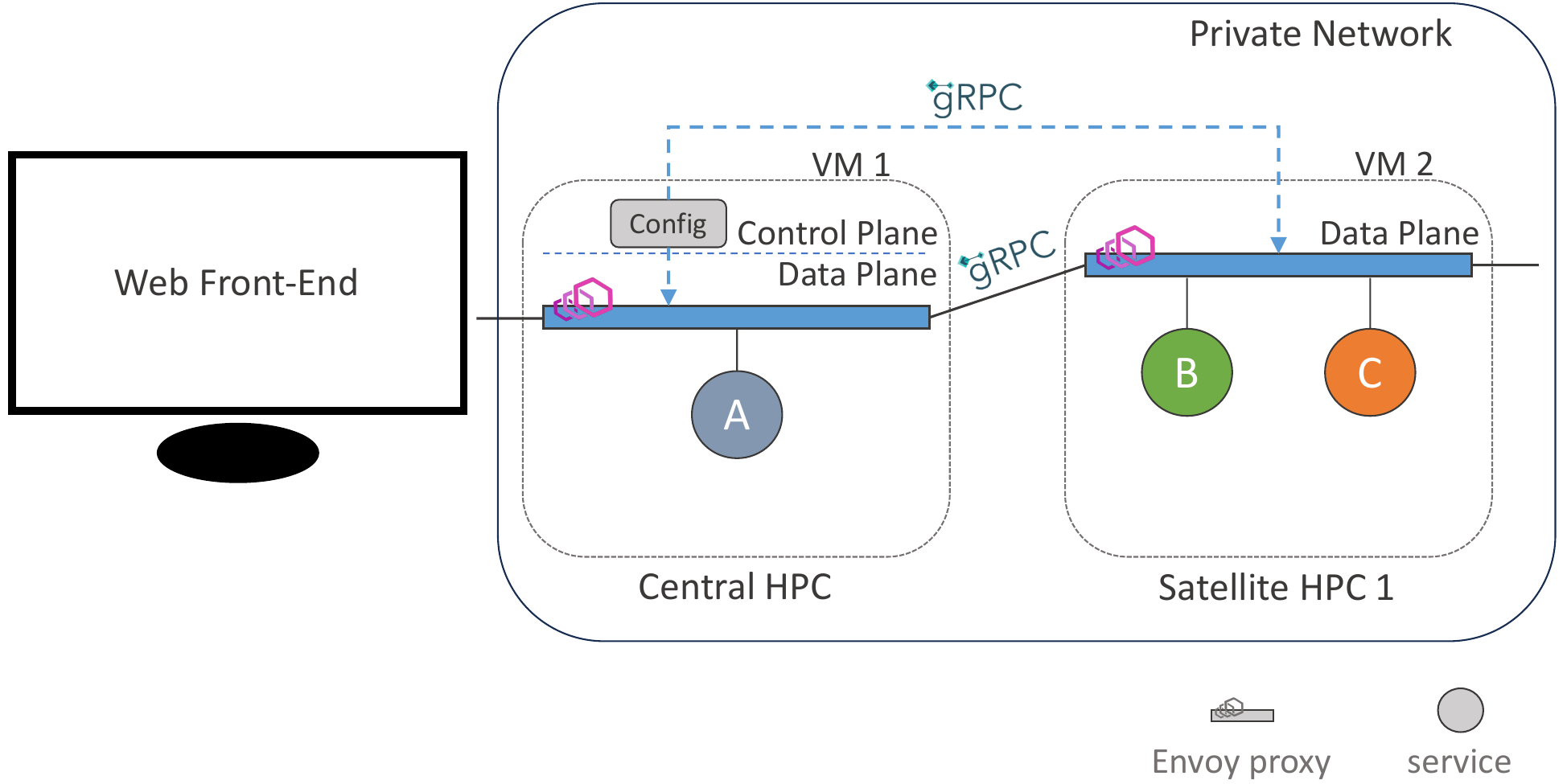}
\caption{Setup used for validating CARISMA. The minimalistic orchestrator has been left out for a better overview. }
\label{fig:val_arch}
\end{figure*}

To simulate the setup within a car, we provisioned two virtual machines that are connected by a private network. Within that private network, every node exposes only the port of its ingress listener to other machines. One of the two virtual machines has been chosen to be the central node and was furthermore connected to a public network in order to expose a web front-end as part of the software component \texttt{A}. That web front-end was intended for testing purposes and displays a value that software component \texttt{A} retrieves from software component \texttt{B}. The other virtual machine was set up as a satellite node. We then deployed (i) - (iv) to the central node and our orchestrator to both nodes. The responsibility of the orchestrator was to deploy the three software components to the corresponding nodes according to a node-specific configuration file. Moreover, the orchestrator was responsible for the registration and deregistration of services upon their deployment and undeployment, respectively. Furthermore, we deployed an instance of Envoy\footnote{Envoy: https://www.Envoyproxy.io} as service proxy per node and connected it to the Control Plane hosted on the central node. We chose Envoy since it is designed with performance and scalability in mind and offers a dynamic configuration interface that we could implement within our Control Plane. Our overall validation setup is depicted in Figure~\ref{fig:val_arch}.

In general, our goal was to validate that two software components that have been deployed based on CARISMA can successfully communicate (\texttt{A -> B}). The third service was employed to validate a proper routing. Concerning the two communicating software components, we validated two scenarios: (i) both applications are running on different nodes, i.e., software component \texttt{B} runs on the satellite node, and (ii) both applications are running on the central node. Moreover, we ensured by validation that switching the deployment target of one of the software components at runtime is possible without a downtime of the other software component. To this end, we instructed the minimalistic orchestrator to deploy the second software component (\texttt{B}) to the satellite node and ensured that the setup works as intended. In a second step, we instructed the minimalistic orchestrator to deploy that software component to the central node as well and, upon successful deployment, remove it from the satellite node. We then ensured that the communication still works without reconfiguring the software component \texttt{A}. 

Overall, our prototype shows the feasibility of the CARISMA approach. It was possible to implement CARISMA with state-of-the-art frameworks, tools, and programming languages. Furthermore, we successfully implemented a distributed application on top of CARISMA and ensured that changing the deployment target of a service at runtime does not affect dependent software components.

\section{\uppercase{Conclusion and Future Work}}
\label{sec:cfw}

In this paper, we present an approach to apply service meshes to the automotive domain. We introduce CARISMA, a service mesh architecture that is specifically designed to be applied to multiple interlinked HPCs. To this end, we differentiate between a central node that combines Control Plane and Data Plane and satellite nodes that are Data Planes only. Furthermore, we require the central node to offer a node and service registry service such that, in a first step, nodes can register with the central node to provide it with their IP address and, in a second step, can register and deregister services when they are deployed or undeployed accordingly. Also, we limit the number of proxies to exactly one per HPC since multiple proxies per HPC would carry the same configuration, and there is no benefit from redundancy at this point. Our overall goal was to enable automotive applications to run distributed across different HPCs without being aware of the actual location of the various involved services. This goal was achieved through our approach. In future research, we will focus on extending CARISMA to further leverage the benefits of service meshes within the automotive domain. Furthermore, we plan to do a sound evaluation of the overhead that CARISMA introduces in comparison to direct inter-service communication.

\section*{\uppercase{Acknowledgments}}

This publication was partially funded by the German Federal Ministry for Economic Affairs and Climate Action (BMWK) as part of the Software-Defined Car (SofDCar) project (19S21002).

\bibliographystyle{apalike}
{\small
\bibliography{ref}}

\end{document}